\documentclass[aps,twocolumn, groupaddress, amsmath,amssymb,nobibnotes,reprint,prb]{revtex4-1}
\usepackage{float} 
\usepackage{graphicx}
\usepackage{dcolumn}
\usepackage{bm}
\usepackage{upgreek}
\usepackage{times}
\usepackage{mathrsfs}
\usepackage{multirow}
\usepackage{hyperref} 
\usepackage{fancyheadings}
\usepackage[version=3]{mhchem}

\begin{document}

\title{Kinetic mechanism for reversible structural transition in \ce{MoTe2} induced by excess charge carriers}

\author{O.~Rubel}
\email[]{rubelo@mcmaster.ca}
\affiliation{Department of Materials Science and Engineering, McMaster University, 1280 Main Street West,
Hamilton, Ontario L8S 4L8, Canada}

\date{\today}

\begin{abstract}
Kinetic of a reversible structural transition between insulating (2H) and metallic (1T') phases in a monolayer \ce{MoTe2} due to an electrostatic doping is studied using first-principle calculations. The driving force for the structural transition is the energy gained by transferring excess electrons from the bottom of the conduction band to lower energy gapless states in the metallic phase as have been noticed in earlier studies. The corresponding structural transformation involves dissociation of \ce{Mo-Te} bonds (one per formula unit), which results in a kinetic energy barrier of 0.83~eV. The transformation involves a consecutive movement of atoms similar to a domain wall motion. The presence of excess charge carriers modifies not only the total energy of the initial and final states, but also lowers an energy of the transition state. An experimentally observed hysteresis in the switching process can be attributed to changes in the kinetic energy barrier due to its dependence on the excess carrier density. 
\end{abstract}


\maketitle

%
%

\section{Introduction}

The quest for materials with resistivity that can be controlled by passing trough a current is driven by their use in data storage and unconventional processing units.\cite{Wang_NM_16_2016,Esser_PNASU_113_2016} Prominent mechanisms realized so far involve formation/dissolution of a conductive filament due to diffusion of ionic species,\cite{Jo_NL_10_2010,Terabe_N_433_2005} or tuning the conductivity via a phase change between crystalline/amorphous states induced by Joule heating.\cite{Wuttig_NM_6_2007} Extreme structural transformations associated with low/high resistance states in those materials naturally limit their endurance. Recently, \citet{Wang_N_550_2017} reported  an experimentally-observed reversible transition between insulating (2H) and metallic (1T') phases in a monolayer \ce{MoTe2} driven by an electrostatic doping. This technique opens up possibilities for developing of new phase-change devices.

\begin{figure}
	\includegraphics{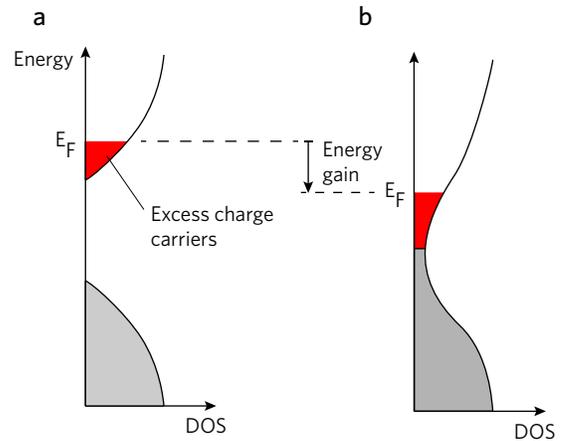}
	\caption{Schematic density of states (DOS) of (a) insulating (2H) and (b) metallic (1T') phases with excess charge carriers. The metallic phase accommodates excess electrons at lower energies. $E_\text{F}$ indicates a position of the Fermi energy.}
	\label{Fig:1}
\end{figure}

The driving force for the 2H$\rightarrow$1T' structural transition in doped transition metal dichalcogenides (TMDs) is the energy gained by transferring excess electrons from the bottom of the conduction band to lower energy gapless states in the metallic phase\cite{Py_CJP_61_1983,Kang_AM_26_2014} as illustrated in Fig.~\ref{Fig:1}. In general, other TMDs with 2H stable structure (e.g., \ce{MoS2}) can undergo a charge-mediated phase transition.\cite{Gao_JPCC_119_2015} However, a high energy difference between the 2H and 1T'  phases (0.8~eV per formula unit (f.u.) of \ce{MoS2}) requires a high excess charge density that is beyond practical capabilities of the electrostatic gating.\cite{Ye_S_338_2012}

The advantage of \ce{MoTe2} is that 2H and 1T' phases are very close in energy. \citet{Li_NC_7_2016} and \citet{Zhang_AN_10_2016} calculated the carrier density required to drive the 2H$\rightarrow$1T' transition in \ce{MoTe2} assuming that the crossover takes place when two phases have the same energy. The threshold excess electron densities of $n_e=3.7\times10^{13}$~cm$^{-2}$ and $6\times10^{13}$~cm$^{-2}$ were predicted by these two groups,\cite{Li_NC_7_2016,Zhang_AN_10_2016} respectively. The experimental values are $1.2\times10^{14}$~cm$^{-2}$ for the 2H$\rightarrow$1T' transition and $5\times10^{13}$~cm$^{-2}$ for the reverse transformation 1T'$\rightarrow$2H.\cite{Wang_N_550_2017} Although the agreement is not perfect, it indicates that first-principle calculations capture the essence of a charge-induced phase transition. What is not addressed so far is a wide hysteresis of $n_e$'s associated with the switching process suggesting a kinetic barrier involved.\cite{Wang_N_550_2017} The goal of this paper is to investigate the energy landscape for 2H$\rightarrow$1T' transformation in \ce{MoTe2} in the presence of an excess charge.

\section{Method}

Vienna \textit{ab-initio} simulation program (VASP) \cite{Kresse_PRB_54_1996, Kresse_CMS_6_1996} density functional theory\cite{Kohn_PR_140_1965} (DFT) package was employed in this work. A meta-generalized gradient approximation SCAN \cite{Sun_PRL_115_2015} with a revised Vydrov-van Voorhis (rVV10) long-range van der Waals interaction\cite{Vydrov_JCP_133_2010,Sabatini_PRB_87_2013,Peng_PRX_6_2016} was used for the exchange-correlation functional since it accurately captures both structural properties and the strength of chemical bonds.\cite{Sun_NC_8_2016} The inclusion of the van der Waals interaction is essential for layered structures.\cite{Bjoerkman_JCP_141_2014} 

Calculations for the bulk \ce{MoTe2} were performed using $10\times10\times2$ and $5\times9\times2$ $k$-mesh for the primitive Brillouin zone of hexagonal and monoclinic phases, respectively. The structural relaxation was performed by minimizing Hellmann-Feynman forces and stresses below 20~meV/{\AA} and 0.5~kbar, respectively. The cutoff energy for the plane-wave expansion was set at 280~eV, which is 25\% higher than the value recommended in the pseudopotential for molybdenum. These parameters ensure better than 10~meV convergence of the total energy difference between 2H and 1T' phases of \ce{MoTe2}. The calculated structural parameters for 2H and 1T' phases are listed in Table~S1 (see Supplementary information) along the side with experimental values.\cite{Puotinen_AC_14_1961,Brown_AC_20_1966} Their good agreement gives a confidence in results of calculations.

The monolayers were derived from the bulk structures with a subsequent relaxation of in-plane structural parameters, while maintaining a fixed spacing $c$ between the layers. The separation of $c=100$~{\AA} was used to represent a monolayer unless otherwise specified.

\section{Results and Discussion}
\begin{figure}
	\includegraphics{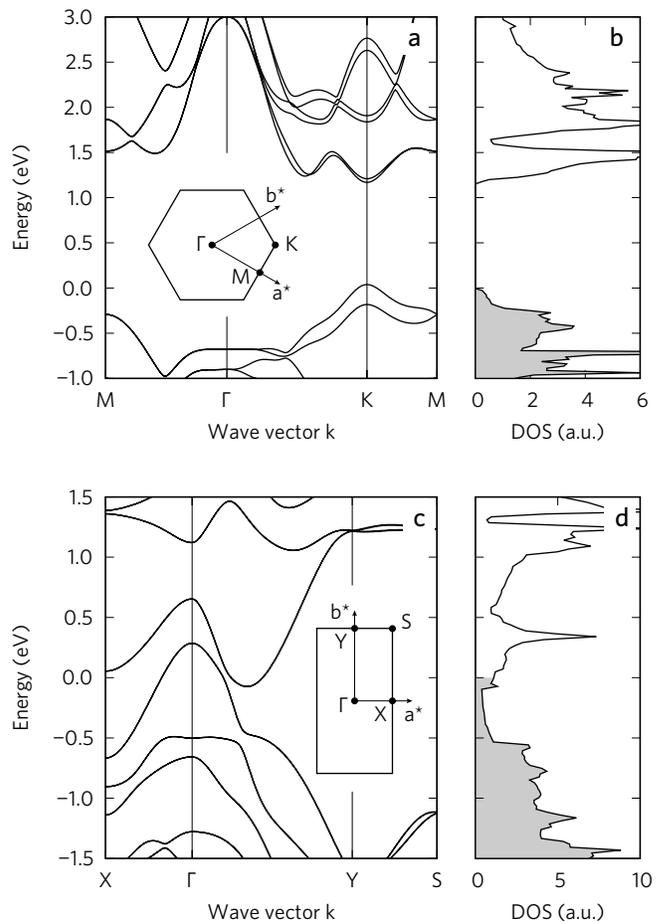}
	\caption{Relativistic band structure and DOS for the monolayer (a,b) 2H-\ce{MoTe2} and (c,d) 1T'-\ce{MoTe2}. The origin of an energy scale is set at the Fermi energy. The corresponding Brillouin zones with high-symmetry points are shown in the insets.}
	\label{Fig:2}
\end{figure}

First we discuss the band structure of the 2H and 1T' phases of \ce{MoTe2} shown in Fig.~\ref{Fig:2}. The monolayer 2H phase features a direct band gap at K~point (Fig.~\ref{Fig:2}a) in accord with optical experiments.\cite{Lezama_NL_15_2015,Froehlicher_PRB_94_2016} The magnitude of the band gap 1.1~eV is in good agreement with the experimental 1.1$-$1.2~eV.\cite{Lezama_NL_15_2015,Froehlicher_PRB_94_2016} The agreement for band gaps is not typical for bare DFT. The likely reason is an error cancelation between an underestimation of the band gap in DFT and a strong excitonic red shift\cite{Ramasubramaniam_PRB_86_2012} present in optical spectra of 2D materials. The bulk 2H structure exhibits a smaller \textit{indirect} band gap (see Fig.~S1 in the Supplementary information), which follows the experimental trends, namely, the transition from a  direct to indirect band gap as the number of layers increases.\cite{Lezama_NL_15_2015,Froehlicher_PRB_94_2016}

The monolayer 1T' phase has a semimetallic band structure (Fig.~\ref{Fig:2}c) with electron and hole pockets  approaching but not touching each other near $\Gamma$ point. In the bulk 1T' phase the electron and hole pockets penetrate each other (see Fig.~S1 in the Supplementary information) giving rise to non-trivial topological states (type-II Weyl semimetal).\cite{Deng_NP_12_2016} This result suggests a few layers \ce{MoTe2} as an alternative candidate for a tunable Weyl fermion metallic state that were previously reported for a Mo$_x$W$_{1-x}$Te$_2$ alloy.\cite{Chang_NC_7_2016}

Calculations of a Fermi energy alignment for a monolayer 2H and 1T' phases of undoped \ce{MoTe2} indicate a validity of the energy argument shown schematically in Fig.~\ref{Fig:1}. Specifically, the Fermi energy of the 1T' phase is located in the middle between highest occupied and lowest unoccupied states of the 2H structure. The total energy of the 2H phase is 40~meV/f.u. lower than that of the 1T' phase in the bulk and only 5~meV/f.u. lower at the monolayer level.

\begin{figure}
	\includegraphics{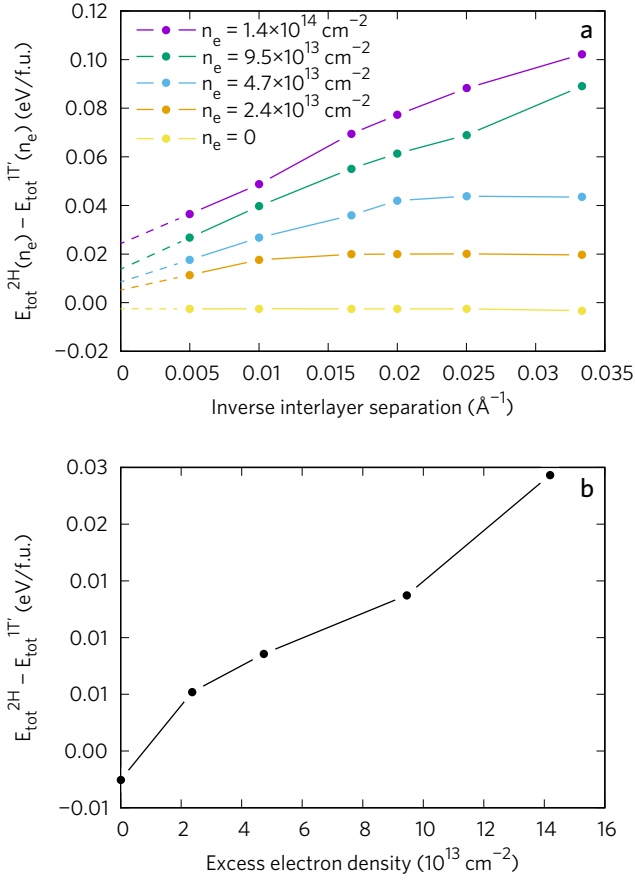}
	\caption{(a) Total energy difference of the monolayers 2H-\ce{MoTe2} and 1T'-\ce{MoTe2} as a function of the inverse interlayer separation $1/c$ shown at various excess electron densities $n_e$. The total energy difference is extrapolated to $1/c=0$ to represent the result for free standing monolayers (dashed lines). (b) Difference on the total energy of the monolayers 2H-\ce{MoTe2} and 1T'-\ce{MoTe2} extrapolated to the limit $c\rightarrow\infty$ as a function of the excess electron density. Without the excess charge the 2H monolayer structure is only slightly more stable than the 1T' alternative ($E_\text{tot}^\text{2H}-E_\text{tot}^\text{1T'}\approx-5$~meV/f.u.).}
	\label{Fig:3}
\end{figure}

Stability of 2H \textit{vs} 1T' phase in the presence of excess electrons of the areal density $n_e$ added to the monolayer \ce{MoTe2} to simulate an electrostatic gating, was evaluated using the total energy difference $E_\text{tot}(n_e)^\text{2H}-E_\text{tot}^\text{1T'}(n_e)$. It is expected that spurious contributions to the total energy of the charged cell will cancel out when subtracting the total energies of two equivalently charged cells. No additional corrections are applied to the charged cell calculations. Results for the energy difference between 2H and 1T' phases are shown in Fig.~\ref{Fig:3}(a) also for different interlayer separations $c$. Data in Fig.~\ref{Fig:3}(a) allow to extrapolate the energy difference to the limit of $1/c=0$, which correspond to a free standing monolayer. The total energy differences presented in Fig.~\ref{Fig:3}(b) indicate that the 1T' phase becomes energetically favorable in the presence of excess electrons.

\begin{figure}
	\includegraphics{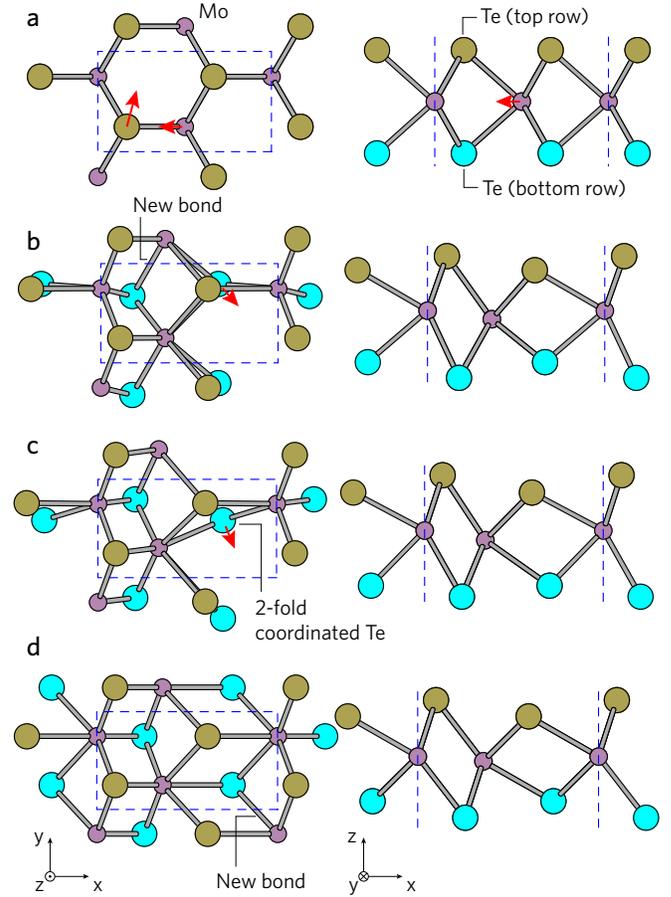}
	\caption{Structural transformation from (a) 2H $\xi=0$, through transitions state (b) $\xi=0.5$ and (c) $\xi=0.7$, to the final state (d) 1T'-phase $\xi=1$ in monolayer \ce{MoTe2}. Red arrows show the direction and magnitude of atomic displacements in the course of the transformation. The transformation is accompanied by dissociation of one \ce{Mo-Te} bond per formula unit.}
	\label{Fig:4}
\end{figure}

\begin{figure}
	\includegraphics{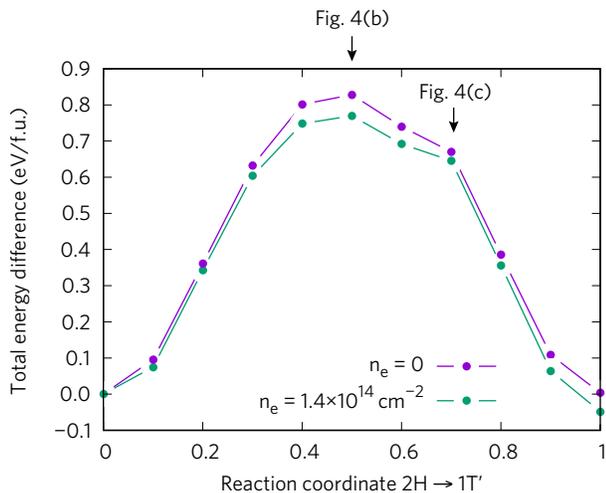}
	\caption{Evolution of the total energy along the 2H$\rightarrow$1T' structural transition paths in the monolayer \ce{MoTe2} calculated without excess charge carriers ($n_e=0$) and with the excess electron density of $n_e=1.4\times10^{14}$~cm$^{-2}$. Arrows indicate two plateaus linked to the corresponding structures in Fig.~\ref{Fig:4}.}
	\label{Fig:5}
\end{figure}

To investigate kinetics of the 2H$\rightarrow$1T' structural transformation, an interpolation formula
\begin{equation}\label{Eq:1}
	\mathbf{a}(\xi)=\mathbf{a}^\text{2H}(1-\xi) + \xi \mathbf{a}^\text{1T'}
\end{equation}
was used to generate intermediate states. Here $\mathbf{a}$ stands for lattice vectors or fractional coordinates of atomic positions, and $\xi$ is the reaction coordinate. Figures~\ref{Fig:4}a and \ref{Fig:4}d illustrate the initial 2H structure ($\xi=0$) and the final 1T' structure ($\xi=1$). A nudged elastic band method\cite{Mills_SS_324_1995} was employed to explore the reaction coordinate space for internal degrees of freedom and find the lowest energy transition state (Fig.~\ref{Fig:5}). An evolution of the total energy as a function of the reaction coordinate exhibits two plateaus that implies a two-step transformation process. The first plateau at $\xi=0.5$ corresponds to the transition state. It  involves contraction of two raw of Mo-atoms along $x$-axis and simultaneous ``squeezing" of the first Te-atom between two Mo-atoms (Fig.~\ref{Fig:4}b). The transformation is accompanied by dissociation of one of \ce{Mo-Te} bonds, which explains a relatively high magnitude of the barrier, 0.83~eV/f.u. The calculated kinetic barrier can be compared to the literature values of 0.9~eV/f.u. for \ce{MoTe2}\cite{Huang_PCCP_18_2016} and 1.6~eV for \ce{MoS2}\cite{Gao_JPCC_119_2015}, which is indicative of a stronger \ce{Mo-S} bond.

The second plateau in the energy landscape occurs near $\xi=0.7$ (Fig.~\ref{Fig:5}). At this point, the second Te-atom transitions between two Mo-atoms (Fig.~\ref{Fig:4}c) and completes the structural transformation. An importance of a correlated movement between Te and Mo atoms during the 2H$\rightarrow$1T' structural transition was emphasized by \citet{Huang_PCCP_18_2016} Furthermore, the transitions steps in Fig.~\ref{Fig:4} show that Te-atoms do not move simultaneously, but rather overcome the barrier in two consecutive steps. As a result, the phase boundary moves from left to right along $x$-axis in Fig.~\ref{Fig:4}. This process is reminiscent of a domain wall motion during polarization switching in ferroelectric materials.\cite{Beckman_PRB_79_2009,Ahmed_MSMSE_22_2014} By analogy, the 2H$\rightarrow$1T' phase transition in \ce{MoTe2} does not happen concurrently, but rather involves propagation of a wave front of a finite width (significantly grater that the simulation cell used here), which requires overcoming a much lower energy barrier. Similar to ferroelectric materials, the 2H$\leftrightarrow$1T' switching process also features a wide hysteresis with respect to the driving force (the excess charge carrier density).\cite{Wang_N_550_2017}

The presence of an excess charge modifies not only the total energy of the final states (Fig.~\ref{Fig:3}), but also affects the energy of a transition state (Fig.~\ref{Fig:5}). The kinetic energy barrier is lowered by $\Delta E_\text{a}=-60$~meV at the doping level of $n_e=1.4\times10^{14}$~cm$^{-2}$. This result can be attributed to the metallic nature of electronic structure at the transition state ($\xi=0.5$ in Fig.~\ref{Fig:5}) and its lower energy in the charge state (Fig.~\ref{Fig:3}) with respect to the 2H structure ($\xi=0$). To estimate the effect of doping on the 2H$\rightarrow$1T' transition rate, we assume that it can be described by the Arrhenius equation $\nu_0\exp(-E_\text{a}/k_\text{B} T)$ with $\nu_0$ being the attempt frequency, $k_\text{B}$ being the Boltzmann's constant, and $T$ being the temperature. A doping-induced modification of the kinetic barrier results in increasing of the transition rate by a factor of $\exp(-\Delta E_\text{a}/k_\text{B} T)$. Here a 10-fold rate increase is expected at the room temperature and the doping level of $n_e=1.4\times10^{14}$~cm$^{-2}$. It should be noted that the electrostatic doping has a similar effect on the energy barrier for the 1T$\leftrightarrow$1T' structural transition in \ce{MoTe2},\cite{Kim_PRB_95_2017} but the energy scale is much smaller (in the order of 1$-$2~meV).

\textbf{Note added during publication:} \citet{Krishnamoorthy_N_10_2018} recently reported a similar study of the kinetic energy barrier and its change due to the presence of electronic excitations on the monolayer \ce{MoTe2}. Their energy barrier of 0.77~eV/f.u. is comparable to 0.83~eV/f.u. reported here in the absence of the excess charge. A qualitatively similar reduction of the energy barrier in the presence of excitation is noticed.

%
%

\section{Conclusion}

The density functional theory with a van der Waals correction was used to study a kinetic barrier in the phase transition between stable insulating (2H) and metastable metallic (1T') phases of \ce{MoTe2}. In bulk, the total energy of 2H phase is 40~meV/f.u. lower than that for the 1T' phase. The energy difference reduces down to 5~meV/f.u. for the monolayer structures indicating that exfoliation stabilizes the 1T' phase. This balance can be further shifted in favor of the 1T' phase by adding excess charge carriers (electrons). The 2H$\rightarrow$1T' structural transformation requires overcoming the energy barrier of 0.83~eV/f.u., which is reduced by 60~meV/f.u. at the excess carrier density of $n_e=1.4\times10^{14}$~cm$^{-2}$ that corresponds to adding 0.15 electrons per formula unit of \ce{MoTe2}. The structural transformation takes place in two steps during which two Te-atoms transit between 2H and 1T' equilibrium positions in a consecutive manner. This finding suggests existence of  commonalities between the 2H$\rightarrow$1T' phase transition in \ce{MoTe2} and a domain wall motion, thereby explaining the presence of a hysteresis in the reversible 2H$\leftrightarrow$1T' phase transition.

%
%
\begin{acknowledgments}
The computations were performed using Compute Canada resources, including an infrastructure funded by the Canada Foundation for Innovation.
\end{acknowledgments}

%
%

\end{document}